# Anisotropic Stick-Slip Behavior of Aqueous Drops on Lubricated Chemically Heterogeneous Slippery Surfaces

Meenaxi Sharma[1], Shivam Gupta[1], Bidisha Bhatt[1], Geeta Bhatt[2], Shantanu Bhattacharya[2], and Krishnacharya Khare[1]*

[1]Department of Physics, Indian Institute of Technology Kanpur, Kanpur – 208016, India

[2]Department of Mechanical Engineering, Indian Institute of Technology Kanpur, Kanpur – 208016, India

*Email: kcharya@iitk.ac.in

**Abstract**

Conventional slippery surfaces show isotropic drop mobility in all directions, but many applications require directional drop motion along a particular path only. In previous studies, researchers used topographic substrates, together with different external stimuli, to demonstrate anisotropic drop motion, which is not very efficient and cost-effective. Herein, we report a novel approach to smartly control drop motion on lubricating fluid coated chemically heterogeneous surfaces composed of alternating hydrophobic and hydrophilic stripes. Upon depositing an aqueous drop on such a surface, the underneath lubricating fluid dewets from the hydrophilic regions but remains intact on the hydrophobic ones, providing sticky and slippery areas for the drop. This results in remarkable anisotropic drop sliding behavior, from uniform motion along parallel to stripes to stick-slip motion along the perpendicular to them. Furthermore, we also demonstrate a phase diagram summarizing different dynamic situations exhibited by drops, sticking, or moving in one or both directions.

# 1. Introduction

The directional motion of liquid drops is very useful in numerous practical applications such as water harvesting, fluid transport, lab-on-a-chip devices, drug delivery, to name a few[1-5]. One can achieve such motion on hydrophobic or superhydrophobic surfaces, but they are not very efficient due to a large amount of pinning sites on dry surfaces[6,7]. Alternatively, inspired by *Nepenthes* pitcher plants, Slippery Liquid Infused Porous surfaces (SLIPs) provide almost frictionless motion to a variety of liquid drops[8-12]. Due to the isotropic nature of underneath roughness (ordered or random) of a solid surface, liquid infused slippery surfaces show uniform drop mobility in all directions[13,14]. Previous studies used anisotropic micro or nano-patterning of the underlying solid surfaces to obtain directional motion of aqueous drops on lubricating fluid infused slippery surfaces[15-19]. Due to the underlying topography, the motion of drops is restricted only along the direction of the fabricated pattern. Recently, researchers demonstrated on-demand reversible liquid transport on lubricating fluid infused slippery surfaces using different external stimuli, for example, electric field, magnetic field, temperature, light, and mechanical strain on patterned lubricated surfaces[12,20-29]. In all these examples, the pattern of underlying topography governs the direction of drop motion, and changing that direction requires repatterning of the surface. Alternatively, chemically heterogeneous surfaces, with the pattern of hydrophilic and hydrophobic regions, have been used in microfluidics for directional liquid transport[30-35]. The major problem associated with chemically heterogeneous surfaces is strong pinning sites for moving drops at chemical heterogeneities, which results in reduced drop mobility on them[36]. Due to the absence of topographic patterns, lubricated chemically heterogeneous slippery surfaces can be an excellent alternative to conventional anisotropic slippery surfaces with improved performance, easy fabrication, and cost-effectiveness. On lubricated chemically heterogeneous surfaces, the biggest challenge is to provide anisotropic lubricated slippery tracks on the underlying chemical pattern, which is the main point of the study of this work.

## 2. Results and Discussion

### 2.1 Stability of lubricating film on LCHet surfaces

We fabricated chemically heterogeneous surfaces, having alternating hydrophilic and hydrophobic stripes with large wettability contrast, using standard microcontact printing (see Supporting Information, Section 1 Figure S1). Figure 1a & b show schematics of equilibrium morphologies of aqueous drops on a dry chemically heterogeneous (CHet) surface, with anisotropic drop shape and two different contact angles ($\theta_\parallel$ and $\theta_\perp$). The underlying pattern of alternating hydrophilic and hydrophobic stripes is responsible for the non-circular three-phase contact line (TPCL) of water drops on it. Figure 1c shows schematics of an aqueous drop deposited on a lubricated chemically heterogeneous (LCHet) surface. To understand the interaction of aqueous drops with dry and lubricated chemically heterogeneous surfaces, rectangular boxes in the inset of Fig. 1b and c show magnified images of the drop-solid interface. On a CHet surface, water wets both the hydrophilic and hydrophobic regions under the drop, and the non-circular TPCL follows the periodic pattern, inward on the hydrophobic and outward on the hydrophilic regions. Whereas on an LCHet surface, a water drop is always surrounded by a wetting ridge formed from the lubricating fluid, and the TPCL remains non-circular and does not lie on the solid surface similar to a homogeneous slippery surface[37-39]. But the most exciting phenomenon happens at the drop-lubricant interface underneath the drop, from where water displaces lubricating oil completely from hydrophilic regions and comes into direct contact with it, whereas it floats on top of lubricating fluid on hydrophobic regions. To understand this phenomenon, we should recall the stability condition that a solid surface should be hydrophobic and oleophilic to use it for stable slippery surface[8,10,40-42]. Whereas if the solid surface is hydrophilic, aqueous drops will sink due to dewetting of the lubricating film underneath the drop. Therefore for an LCHet surface, the lubricating fluid from the hydrophilic region underneath an aqueous drop becomes unstable and dewets, as a result, water comes into direct contact with the hydrophilic region of the solid surface. On the hydrophobic region, the lubricating fluid underneath the drop remains intact, hence the part of water drop on top of lubricating oil above hydrophobic region floats. This configuration is shown schematically in the rectangular box in Fig. 1c and should be the lowest energy configuration compared to other possible configurations.

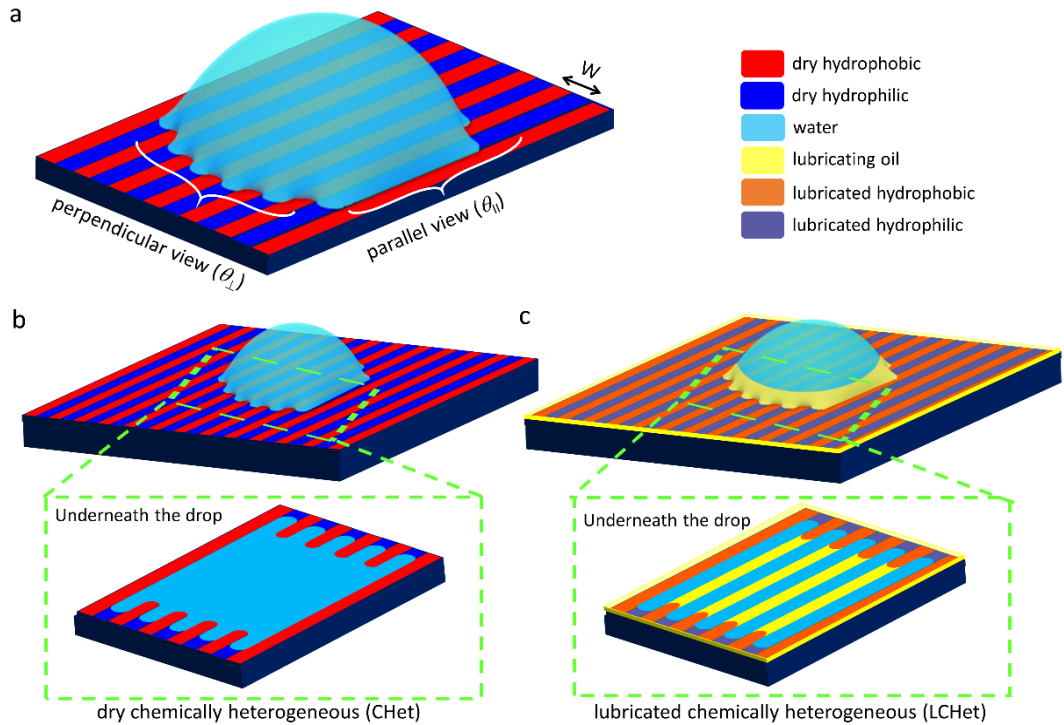

**Figure 1: Schematics of water drops on CHet and LCHet surfaces. a,** 3-dimensional drop shape on a dry chemically heterogeneous surface showing parallel ($\theta_\parallel$) and perpendicular ($\theta_\perp$) contact angles. **b, c,** Drop shapes on CHet and LCHet surfaces respectively with magnified images of underneath the drops shown in the rectangular boxes. On CHet surfaces, water wets both the stripes, whereas on LCHet surfaces, water wets only the hydrophilic stripes as the lubricating oil dewets from hydrophilic regions. On hydrophobic regions, a thin lubricating oil film is present between water and the surface.

Following the energy minimization approach, one can derive the stability criteria of lubricating fluid on LCHet surfaces as equations (1) & (2) in a way similar to homogeneous lubricated surfaces (see Supporting Information, Section 2 for detailed derivation).

$$\Delta E_1 = \gamma_{OV} \cos\theta_{OS_2} - \gamma_{WV} \cos\theta_{WS_2} - \gamma_{OW} < 0 \qquad (1)$$

$$\Delta E_2 = \gamma_{WV} \cos\theta_{WS_1} - \gamma_{OV} \cos\theta_{OS_1} + \gamma_{OW} < 0 \qquad (2)$$

where $\Delta E_1$ and $\Delta E_2$ are energy differences between the state corresponding to Fig. 1c and two other possible states. $\gamma$ is the interfacial tension, and $O, V, W,$ and $S$ represent oil, vapor, water, and solid phases respectively, and subscripts 1 and 2 correspond to hydrophobic and hydrophilic regions, respectively. Therefore, LCHet surfaces provide an anisotropic pattern of the lubricating fluid in the form of lubricant microchannels, which can demonstrate anisotropic slippery behavior.

To control the anisotropic slippery behavior, we prepared various chemically heterogeneous samples with different area fraction of hydrophobic and hydrophilic regions (see Supporting Information, Section 3 Table S1). The samples were characterized using water condensation and phase imaging with an atomic force microscope (see Supporting Information, Section 4 Figure S3). We investigated that the static wetting behavior of aqueous drops on LCHet surfaces (measured as apparent contact angle in the parallel direction ($\theta_{app(\parallel)}$)), is very different compared to CHet surfaces ($\theta_{\parallel}$) (see Supporting Information, Section 5 Figure S4). This difference is mainly due to the instability of lubricating fluid film underneath the drops on LCHet surfaces, which is shown in Fig. 2a. When water drops are deposited on LCHet surfaces with uniform hydrophilic ($S_0$) or hydrophobic ($S_8$) surface wettability (cf. Fig. 2a i and iv), the drops sink completely or float on them respectively, as predicted by the sinking/floating condition discussed earlier. Due to the instability of the lubricating film underneath a water drop, the apparent contact angle of the water drops on samples $S_0$ decreased rapidly from 109° to 52° in just 5 s, which later saturated to about 48°. The initial sharp decrease in the apparent contact angle is due to the dewetting of the underneath lubricating film, bringing water drops into direct contact with the hydrophilic solid surface. Fluorescence image of sample $S_0$ in Fig. 2c shows the final dewetted oil droplets underneath the water drop, confirming the direct contact of water drops with the hydrophilic solid surface. On the other hand, on samples $S_8$ having a hydrophobic solid surface, the apparent contact angle of a water drop did not change much, suggesting a negligible change in the status of underneath lubricating oil film. Fluorescence image of sample $S_8$ in Fig. 2c shows a uniform red color confirming the presence of uniform silicone oil film underneath the water drop. However, on LCHet surfaces $S_1$ & $S_5$, where both wettability regions are present, only partial sinking of water drops are observed (cf. Fig. 2a ii and iii), which is evident from the intermediate values of the apparent contact angles. Fluorescence image of sample $S_1$ in Fig. 2c confirms the complete removal of lubricating oil from hydrophilic regions, whereas it remains intact on the hydrophobic regions (see Supporting Information, Movie S1).

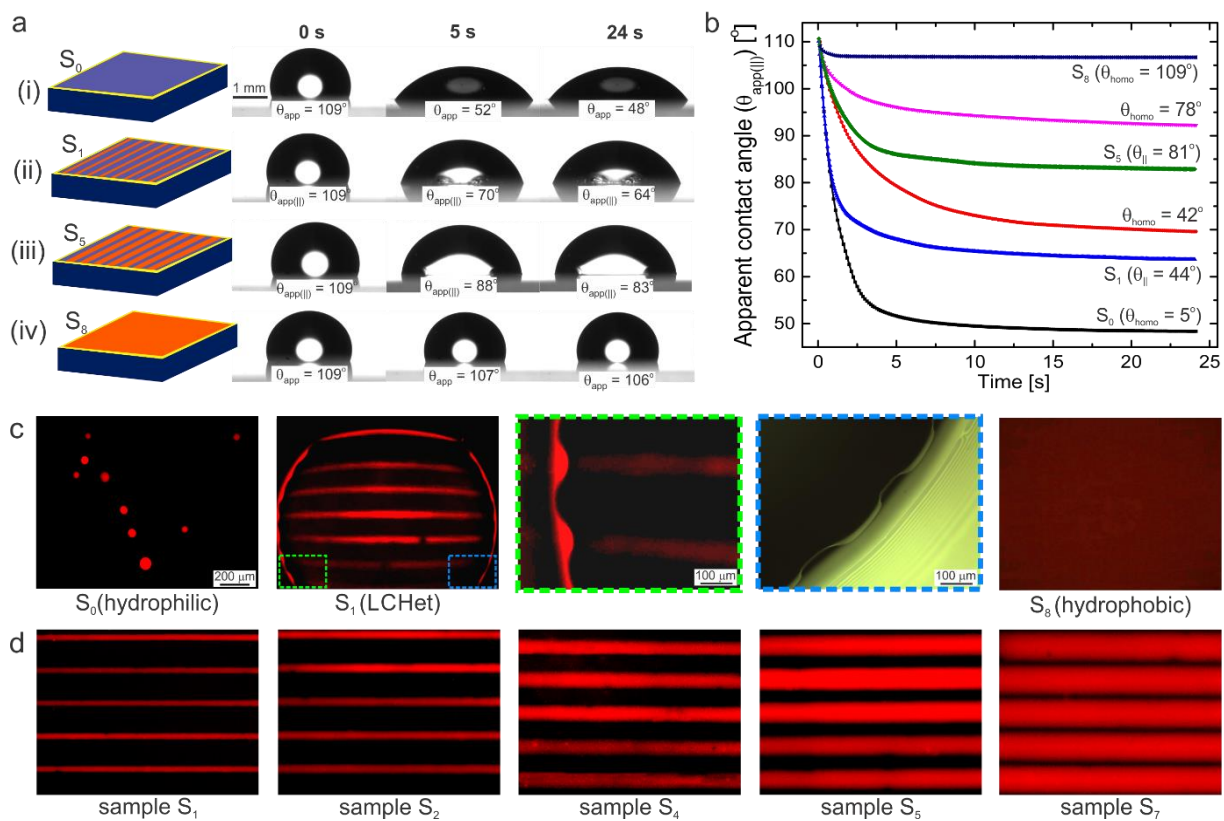

**Figure 2: Stability of lubricating film on homogenous and heterogeneous lubricated surfaces. a,** Snapshots of sinking water drops on (i) lubricated homogenous hydrophilic ($S_0$) ($\theta_{homo}$ = 5°) (ii) LCHet $S_1$ (iii) LCHet $S_5$ and (iv) lubricated homogenous hydrophobic ($S_8$) ($\theta_{homo}$ = 109°) surfaces. **b,** Comparing sink dynamics on lubricated homogenous and heterogeneous surfaces with almost similar apparent contact angles on dry surfaces. **c,** Fluorescence optical images of lubricating oil film on samples $S_0$, $S_1$, and $S_8$ showing dewetted oil drops, dewetting oil microchannels, and stable oil film, respectively. Magnified image of sample $S_1$ in the green box shows a non-circular three-phase contact line that follows the wettability pattern of the solid surface, and in the blue box shows a fringe pattern near the TPCL due to the presence of a wetting ridge around the drop. **d,** Fluorescence images of dewetted oil pattern on samples $S_1$-$S_7$, with increasing hydrophobic area

fraction, showing increasing width of lubricant microchannels. Scale bar for all figures is the same as 200 μm.

One can even see the reflection of dewetted lubricating film from hydrophilic regions while measuring the apparent contact angle in the perpendicular direction ($\theta_{app(\perp)}$) on LCHet surfaces (see Supporting Information, Section 5 Figure S4). Experimentally observed states of lubricating fluid on LCHet surfaces (i.e. dewetting from hydrophilic regions and formation of lubricant microchannels on hydrophobic regions) can also be verified using the short- and long-range forces (in terms of Spreading coefficient and Hamaker constant respectively) acting on the system (see Supporting Information, Section 6). It is interesting to note here that lubricant gets completely displaced from the hydrophilic regions of sample $S_1$, however on sample $S_0$ lubricant forms multiple micron-sized droplets after complete dewetting. This is because, on LCHet surfaces, adjacent hydrophobic regions act as a reservoir for the dewetted lubricating fluid so the lubricant can easily be entirely displaced from the hydrophilic regions to the hydrophobic regions. But for a homogenous hydrophilic surface, the three-phase contact line of a water drop acts as a barrier between the lubricant outside and underneath the drop. Therefore the underneath lubricating film dewets like a thin film forming multiple droplets after complete dewetting. Figure 2c also shows the magnified images of the area close to the TPCL of a water drop on sample $S_1$, where the green box shows non-circular TPCL following the wettability pattern and presence of lubricating oil only on the hydrophobic regions. The blue box shows interference fringes close to TPCL, indicating the increase in lubricating oil thickness at the TPCL due to the presence of wetting ridge. Sample $S_1$ with larger area fraction of hydrophilic region, showed more sinking (smaller value of $\theta_{app(\parallel)}$) compared to samples $S_5$ having smaller area fraction of hydrophilic region.

  Subsequently, we also compared the sinking behavior of water drops on LCHet surfaces with uniform wettability ($\theta_{homo}$) lubricated surfaces of similar contact angles (cf. Fig. 2b). Since both the uniform wettability surfaces ($\theta_{homo}$ = 42° & 78°) used in the experiments were hydrophilic, the lubricating films would always dewet upon depositing water drops on them. However, the final dewetting pattern would be different as it depends strongly on the surface wettability. It is clear from Fig. 2c that the sinking behavior on both the surfaces is quite similar, however uniform wettability ($\theta_{homo}$) lubricated surfaces always show less sinking for water drops compared to the corresponding LCHet surfaces. The reason behind this is that water drops are exposed to completely hydrophilic regions (having contact angle 5°) upon dewetting on LCHet surfaces compared to moderately hydrophilic solid surfaces of uniform wettability ($\theta_{homo}$ = 42° & 78°) after dewetting of the lubricating film. Also, the dynamics of sinking on LCHet surfaces are always faster compared to uniform wettability lubricated surfaces for the same reason, as mentioned above. To clearly visualize the dewetted lubricant microchannels in a large area, we repeated similar experiments under a water bath rather than water drops. Figure 2d shows fluorescence images of dewetted lubricated oil pattern on LCHet surfaces $S_1$-$S_7$ showing increasing width of lubricant microchannels with increasing hydrophobic area fraction. Hence we learn that the extent of sinking increases with increasing hydrophilic area fraction of LCHet surfaces, and the wettability of the solid surface plays the most crucial role for uniform wettability as well as chemically patterned lubricated surfaces.

**2.2 Motion of water drops on LCHet surfaces**

Figure 3a shows the rear and front contact points of a moving drop and Fig. 3b shows schematics of drop motion in parallel and perpendicular directions. As we deposit a water drop on a LCHet surface, only the lubricating oil film underneath the drop dewets spontaneously, and the area outside the drop remains uniformly lubricated. Now, as the drop moves on a tilted LCHet surface in either direction due to gravity, the new underneath drop area undergoes dewetting, whereas the area left behind the drop becomes uniformly lubricated due to the flow of the oil. During experiments, we observed that water drops did not move at all in either direction on samples $S_1 - S_4$, i.e., from 20% to 49 % hydrophobic area fractions. On other samples (with hydrophobic area fraction ≥ 65%), water drops were mobile in one or both directions depending on the hydrophobic area fraction. On samples $S_5$ and $S_6$, with 65% and 80% hydrophobic area fractions respectively, water drops moved only along the parallel direction, and only on the sample $S_7$, having 92% hydrophobic area fraction, water drops moved along both the directions (see supporting information Movie S2-Movie S8). It is interesting to note that despite complete dewetting of the lubricating film underneath a drop, which brings a part of the drop in direct contact with the hydrophilic solid surface, the drop can still move on them.

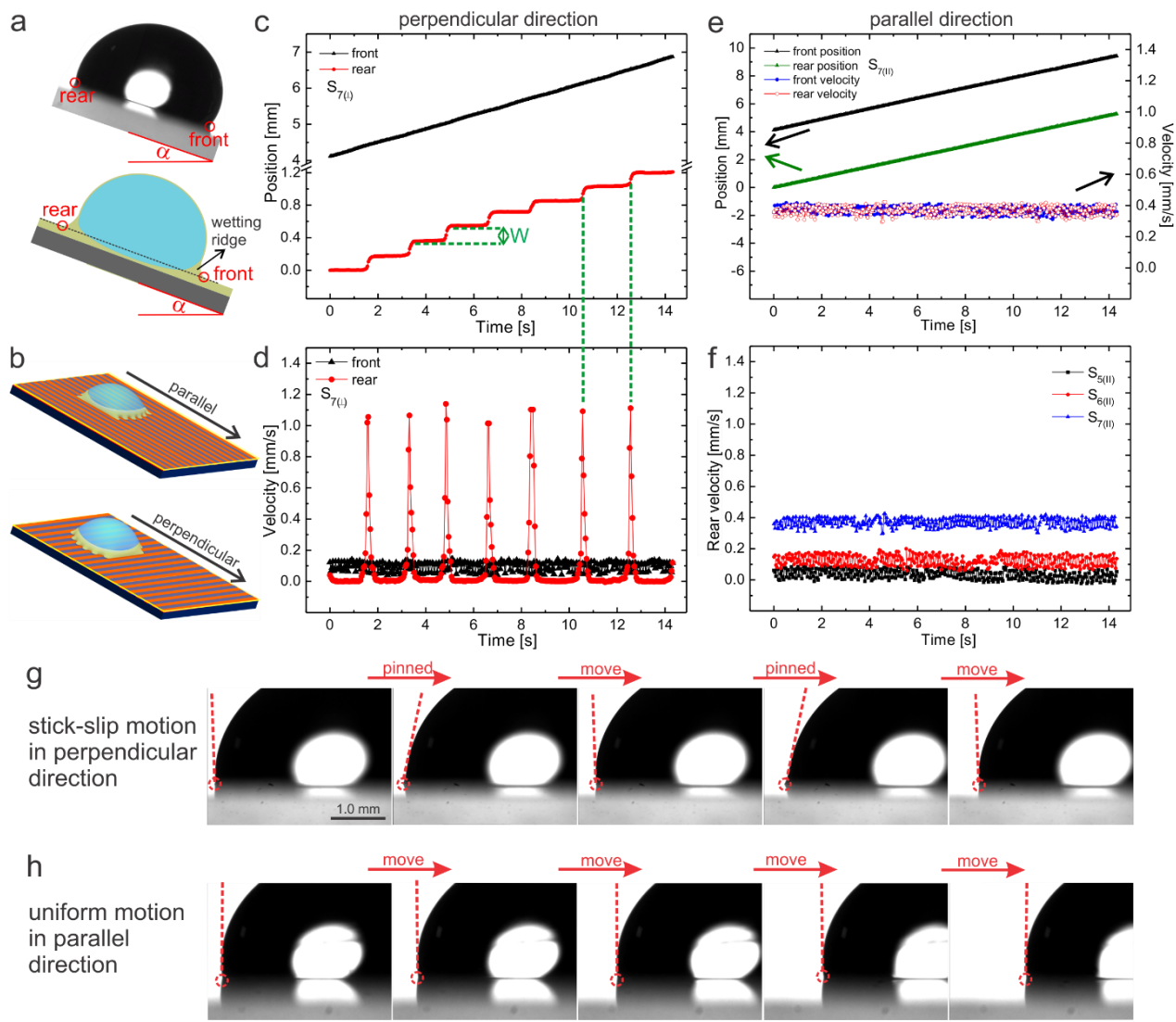

**Figure 3: Drop motion on LCHet surfaces. a,** Demonstration of the front and rear contact points through an optical image and schematic. **b,** Schematic illustration of the drop motion along parallel and perpendicular directions. **c, d,** Positions and corresponding velocities of the front and rear contact points of a moving drop in the perpendicular direction on the sample $S_7$. The front contact point performs smooth motion, whereas the rear contact point exhibits stick-slip motion. **e,** Motion

of a water drop in the parallel direction showing uniform motion for both the front and the rear contact points on the same sample $S_7$. **f,** Velocity of the rear contact points of water drops in the parallel direction for three different surfaces, $S_5$, $S_6$, $S_7$. **g, h,** Optical images of moving water drops on sample $S_7$ along perpendicular and parallel directions showing stick-slip and uniform motion, respectively. Scale bar for all optical images is the same as 1 mm.

The reason behind the motion is the formation of slippery lubricant microchannels on the hydrophobic regions of LCHet surfaces, which helps in overcoming the pinning due to hydrophilic surface. For a drop motion along the perpendicular direction, the drop has to overcome the pinning barrier at every boundary of hydrophilic-hydrophobic region, which is not the case for the motion along the parallel direction. This is the reason that water drops can move along parallel direction even on smaller hydrophobic area fraction samples, whereas perpendicular motion is seen only for one sample with the highest hydrophobic area fraction. Figure 3c and d shows positions and velocities, respectively, of front and rear contact points of a moving drop in the perpendicular direction on a sample $S_7$. It is clear from the figure that the position of the front contact point changes continuously, whereas the position of the back contact point changes in a stepwise manner. This is due to the same reason that the front of the drop faces uniformly lubricated surface, whereas the back contact point experiences dewetted lubricated surface. As a result, the back contact point slips on the lubricated part on the hydrophobic strip but remains pinned on the hydrophilic part from where the lubricant has dewetted. Therefore the gap between two consecutive steps of the back contact position corresponds to the pitch ($W$) of the wettability pattern. Due to the uniform motion, the velocity of the front contact point is constant, whereas the back contact point shows stick-slip velocity pattern, as shown in Fig. 3d. When the contact line is pinned, the drop velocity remains zero which spontaneously shoots up as the contact line depins from the hydrophilic region and slips on the hydrophobic region which again becomes zero at the next hydrophilic boundary and the process similarly repeats itself (see supporting information Movie S2 and Movie S3). When water drops move in the parallel direction, both front and back contact points move continuously with constant velocity as shown in Fig. 3e. This is because, during motion along the parallel direction, moving drops do not cross the wettability pattern or the barrier as they move parallel to them (see supporting information Movie S4). Figure 3f compares the velocity of rear contact points of moving water drops in parallel direction on samples $S_5$, $S_6$, and $S_7$. Since the hydrophobic area fraction increases from sample $S_5$ to $S_7$, the parallel velocity also increases in the same manner (see supporting information Movie S4, Movie S6, and Movie S8). Figure 3g and h show the snapshots of moving drops on sample $S_7$ along perpendicular and parallel directions illustrating stick-slip and uniform motion, respectively. During the stick-slip motion, the TPCL of a moving drop changes from pinned state to moving state, which is indicated by the changing apparent contact angles. Whereas for the uniform motion along the parallel direction, the TPCL moves freely as suggested by constant apparent contact angles.

**2.3 Phase diagram**

Based on the apparent contact angles and drop velocities on LCHet surfaces with increasing hydrophobic area fraction, we also prepared a phase diagram for better illustration of anisotropic wetting behavior and drop motion. Figure 4a shows the phase diagram where the left and right Y-axes correspond to apparent contact angle and drop velocity, respectively, and the X-axis shows percentage hydrophobic area fraction. The horizontal dotted line represents the boundary between apparent hydrophilic and hydrophobic states at an apparent contact angle of 90º. Up to hydrophobic area fraction of 49 %, water drops do not move since both parallel and perpendicular apparent contact angles are less than 90º. Therefore the first dotted vertical line at 49 % hydrophobic area fraction represents the state up to which no drop motion in any direction is observed, hence designated as "sticking" region. Beyond this particular hydrophobic area fraction, a transition from sticking to motion is observed where drops can move only in one direction, i.e. parallel to the stripes. Drop motion in this region is due to the fact that the perpendicular contact angle becomes greater than 90°. As a result, drops can move in the parallel direction, and the region is designated as "only parallel motion" region. This region extends up to 80% hydrophobic area fraction at which the second vertical dotted line is drawn. Further increasing the hydrophobic area fraction results in both the apparent contact angles values greater than 90° indicating the drop motion in both the directions.

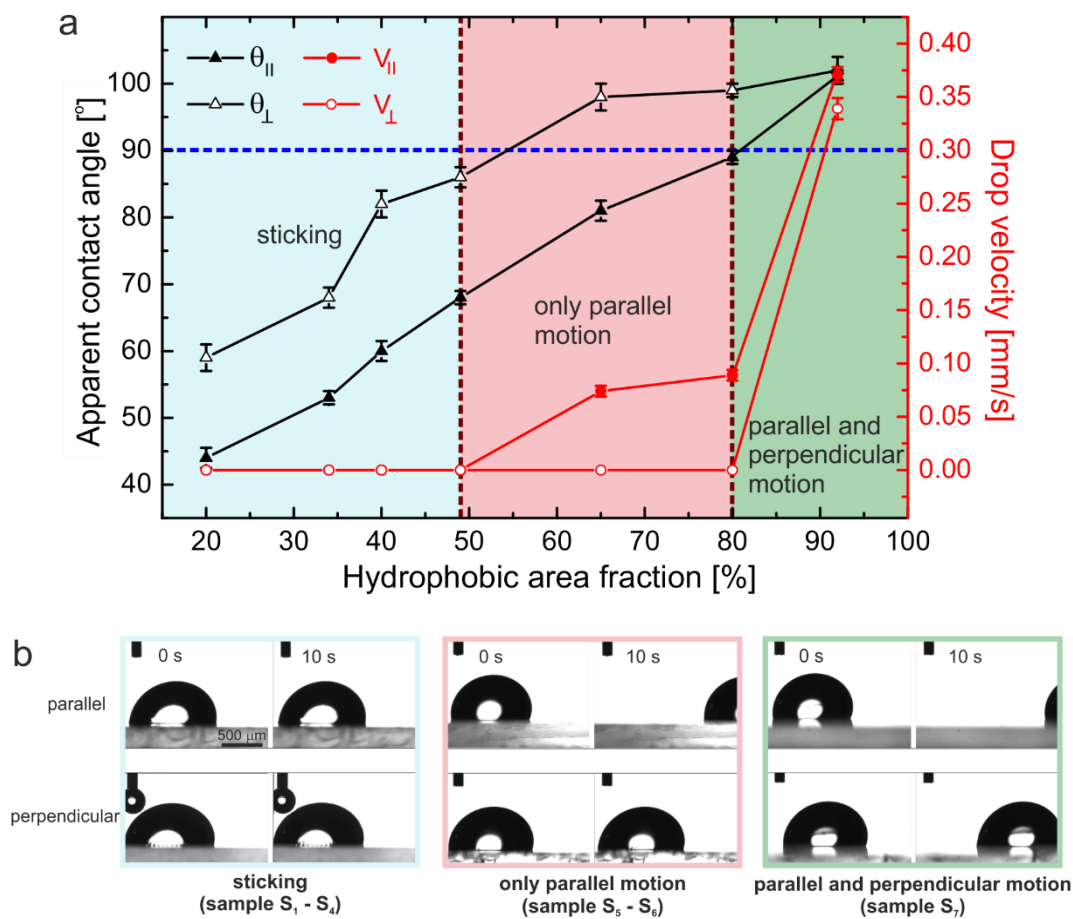

**Figure 4: Phase diagram showing apparent contact angle and drop velocity on LCHet surfaces. a,** Phase diagram showing apparent contact angle and drop velocity of water drops on LCHet surfaces. Based on the apparent contact angle and drop velocity values, the plot can be divided into three regions namely, "sticking" (blue), "only parallel motion" (red) and "parallel and perpendicular motion" (green). Errors correspond to measurements from three different spots on three different samples. **b,** Optical images of moving drops along parallel and perpendicular directions corresponding to the three regions.

Therefore, the third region is designated as "parallel and perpendicular motion" region showing large velocities along both directions. Optical images in Fig. 4b show parallel and perpendicular motion of water drops corresponding to the three regions of the phase diagram. Unfortunately, we could get only one such data point where motion occurred in both the directions, stick-slip in perpendicular and uniform in parallel, due to experimental limitations. But it is clear from the phase diagram that both apparent contact angle should be larger than 90° to observe motion along parallel and perpendicular directions. The above results demonstrate that one can exploit the dewetting of a lubricating oil film from the hydrophilic region of chemically heterogeneous surfaces to obtain lubricating microchannels, which can subsequently be used to illustrate anisotropic drop motion.

**3. Conclusion**

Using the knowledge of dewetting of a thin liquid film and lubricating fluid based slippery surfaces, we report anisotropic slippery surfaces with controlled drop mobility in the desired direction. We achieved this using chemically heterogeneous surfaces, having hydrophobic and hydrophilic stripes, coated with a suitable lubricating fluid. Upon depositing a water drop on them, the lubricating fluid dewets from the hydrophilic region and moves to the neighboring hydrophobic region to minimize the total energy. This results in the formation of lubricant microchannels, which subsequently demonstrate the anisotropic slippery behavior. Drops always move with uniform velocity in the parallel direction, whereas they demonstrate stick-slip motion in the perpendicular direction. Therefore, the magnitude and direction of drop mobility depend on the area fraction of the hydrophobic region and the wettability contrast between the hydrophilic and hydrophobic regions. We also prepared a phase diagram to summarize the anisotropic drop mobility on such lubricated chemically heterogeneous surfaces. We envision that current findings, including the general principles, anisotropic wetting, and drop mobility will be helpful in the use of the anisotropic slippery surfaces in many practical applications.

**4. Experimental section**

Materials: Microscopic glass slides (root mean square surface roughness of 8 (± 2) nm) with dimension 2.5" x 2.5" were used as solid substrates. The glass slides were cleaned using deionized water (Milli-Q Millipore), ethanol (Merck), and acetone (Sigma Aldrich). Polydimethylsiloxane (PDMS) base and curing agent (Sylgard 184) was purchased from Dow Corning, USA. Hexane, Trichloro(octadecyl)silane (OTS) (90%), and Nile Red dye were purchased from Sigma Aldrich. All chemicals were used without any further modification. SU-8 photoresist (MicroChem, Germany) was used to prepare linear channels using photolithography on a glass slide. Commercially available 350 cSt silicone oil (Gelest) was used as a lubricating fluid.

Preparation of CHet and LCHet surfaces: Various steps involved in the preparation of CHet surfaces include mask designing, master surface preparation using standard photolithography and microcontact printing (μCP) are shown in the Supporting Information Figure S1. SU-8 2015 photoresist was spin-coated onto the cleaned glass slides at 3500 rpm and subsequently the substrates were soft baked at 95 °C for 4 min and slowly cooled down for 1 hr. The photoresist coating was then exposed to UV light (365 nm, 40 mW/cm$^2$) using a photomask aligner and post baked at 95 °C for 4 min and finally developed with a SU-8 developer (MicroChem, Germany). The resulting linear channels were 12 μm deep with lateral dimensions (width) varying from 20 μm - 185 μm. Surfaces were then replicated using PDMS in order not only to damage the master surfaces but also to get the desired dimensions since reverse features were formed on the master surface using negative photoresist. These microstructured PDMS replica sheets were bonded on glass slides (backside of PDMS sheet in contact with glass) to use it as a stamp for the microcontact printing. A flat PDMS sheet was soaked in 0.4 wt % solution of OTS in hexane and dried using a nitrogen gun for 20 s. Flat PDMS sheet was placed on a PDMS replica sheet for 6 s to transfer the OTS molecules only at the top of the channels. Subsequently, the PDMS replica sheet was placed in contact with the plasma cleaned glass substrate with a load of 100 gm on top of it. After gently removing the PDMS replica sheet, the glass substrates were annealed at 90 °C for 20 s and rinsed with hexane to remove any non-reacted OTS. To prepare LCHet surfaces, 350 cSt silicone oil (density 970 kg/m$^3$ and surface tension 21 mN/m) was spin-coated at 8000 rpm on the dry CHet surfaces. Resulting LCHet surfaces exhibit an oil film thickness of 4 μm. For fluorescence imaging, 0.2 wt % Nile red dye mixed in a solution of 2 wt % toluene in 350 cSt silicone oil was spin-coated onto the substrate at the same rpm. After spin coating, measurements were performed after half an hour to ensure the complete evaporation of the solvent.

<u>Instruments and Characterization:</u> An optical microscope (BX-51, Olympus) was used to characterize the CHet surfaces by analyzing them through vapor condensation. Lubricating film thickness was estimated using the weight difference method. Fluorescence microscopy was used to investigate the lubricating film underneath water drops. Static wetting behavior of CHet and LCHet surfaces was characterized using water drops of volume 4 μl, which covers approximately 10 to 15 wettability stripes. For drop velocity measurements, water drops of 20 μl volume was gently deposited using a micropipette (Tarson) on the substrates tilted at 15°. All contact angle and drop velocity measurements were performed using the dynamic automatic contact angle goniometer (OCA-35 DataPhysics, Germany). All the snapshots and videos were recorded using a high-speed camera (200 fps) equipped with the goniometer. All measurements were repeated at three different positions on three different samples to calculate the standard deviation.

**Acknowledgments**

The authors would like to acknowledge the funding support from SERB, New Delhi (Project No. CRG/2019/000915) and DST, New Delhi, through its Unit of Excellence on Soft Nanofabrication at IIT Kanpur.

**Author contributions**

K.K. conceived the project, K.K. and M.S. designed the project, M.S., G.B. and S.B. prepared samples, M.S., S.G. and B.B. performed experiments and analyses, M.S. and K.K. wrote the manuscript with inputs from all other authors.

**Competing financial interests**

The authors declare no competing financial interests.